\documentstyle[aps,psfig]{revtex}
\newcommand{\be}{\begin{equation}}
\newcommand{\ee}{\end{equation}}

\newcommand{\psib}{\overline{\psi}}

\newcommand{\dd}{\partial\hspace{-7pt}/}

\newcommand{\ber}{\begin{eqnarray}}
\newcommand{\eer}{\end{eqnarray}}

\begin{document}
\title {\bf QCD phase transition in rotaing neutron star, Neutrino beaming and
Gamma-ray bursters}
\vskip 0.2in
\author{Abhijit Bhattacharyya,$^a$\footnote
{E-Mail :bhattacharyyaabhijit\_10@yahoo.co.uk} Sanjay K. Ghosh$^b$\footnote
{E-Mail : sanjay@bosemain.boseinst.ac.in} and Sibaji Raha$^b$ \footnote 
{E-Mail : sibaji@bosemain.boseinst.ac.in} }
\address{$^a$ Department of Physics, Scottish Church College, 1 \& 3, Urquhart 
Square, Kolkata - 700 006, INDIA}
\address{$^b$ Department of Physics and Centre for Astroparticle Physics 
\& Space Science, Bose Institute, 93/1, A.P.C. Road,
Kolkata - 700 009, INDIA}
\maketitle
\vskip 0.4in
\begin{abstract}
We have studied the emission of neutrinos from a rotating hybrid star. We 
find that the emission is predominantly confined to a very small angle,  
provided the core of the star is in a mixed phase of quarks and hadrons 
and the size of such a mixed phase is small. Annihilation of neutrinos 
to produce gamma rays has been discussed. The estimated duration of the 
burst is found to be within the observational range. 
\end{abstract}
\vskip 0.3in

The existence of strange quark matter (SQM) containing u, d and s quark 
had been postulated quite some time ago. It was also proposed that the 
SQM could be the {\it true ground state} of quantum chromodynamics 
\cite{witten}. This conjecture has been supported by various model 
calculations for certain ranges of the model parameters\cite{bag1,ccdm}. 
If this is so, then the usual hadronic matter could undergo a phase 
transition to SQM at high temperature and/or density. This opens up
the possibility that the interior of neutron stars may consist of SQM, 
the baryon density there being extremely large (8-10 times that of 
nuclear matter density at saturation).

In SQM, the strangeness fraction {\it i.e.} the ratio of strange quark and 
baryon number densities, will be unity, if one considers the masses of u, 
d and s to be same. Even for realistic strange quark masses 
( \( m_s > m_u = m_d \) ), the strangeness fraction in quark matter at 
high temperature/density is not much smaller than unity. In contrast to 
quark matter, the strangeness fraction in hadronic matter is usually found 
to be small \cite{wiringa,glendenning,ghosh1}. Sizable strangeness fraction 
can be accommodated in hadronic models by including hyperons and/or kaon 
condensation \cite{our}. For the mean field models with hyperons, the 
strangeness fraction has been shown to be smaller than unity \cite{ghosh1}, 
at densities where the transition to quark matter may occur. Depending on 
the model parameters and the interactions considered, the situation may be 
different for an equation of state (EOS) with kaon condensation 
\cite{our,thorsson}. But for such cases, the transition to quark matter is 
found to be pushed towards much higher densities.

The above discussion implies that the transition from hadronic matter to 
SQM inside neutron stars may be associated with a large production of 
strangeness. This can also be explained in the following way. Initially, 
hadronic matter, in terms of quark content, consists predominantly of u 
and d quarks and possibly a very small fraction of s quarks (from hyperons). 
The quark matter thus formed through a phase transition from the hadronic 
matter should be out of chemical equilibrium. The weak interactions then 
convert this chemically non-equilibrated matter to equilibrated SQM with 
roughly equal numbers of u, d and s quarks. This conversion is associated 
with the production of large amounts of energy in the form of neutrinos, 
with average energy of the order of 100 MeV \cite{sannpa}. 

The observed Gamma Ray Bursters (GRBs) \cite{grbrev} continue to be a puzzling 
astrophysical phenomenon, insofar as there does not exist an universally 
accepted explanation as yet. In recent times, however, it has been 
argued that most GRBs are associated with 
supernovae \cite{grbrev} and thus, efforts are on to arrive at some 
picture for the GRB engine consistent with the supernova explosion. 
In ref.\cite{bombaci1}, Berezhiani {\it et al.} have suggested that 
the origin of the GRBs may be associated with the deconfinement 
transition inside a neutron star resulting into a hybrid star 
(neutron star with quark matter in its centre, either in a mixed 
phase or in a pure quark matter state) or a pure quark star. 
In their model, the time delay between the supernova explosion 
which creates the neutron star and the GRB is governed by  
matter accretion on the neutron star; only when sufficient matter 
has been accreted does the neutron star core density cross the threshold 
for the phase transition to quark matter. While this picture provides a 
plausible scenario for the GRB engine and explains the time delay between 
the supernova explosion and the GRB, several features of GRB, namely the 
amount of energy release, the duration of the burst and most importantly, 
the observed beaming \cite{frail}, remain unaddressed.

In the present letter, we extend the basic premise of ref.\cite{bombaci1} 
to rotating stars. As a rotating neutron star gets 
converted into a hybrid star, with the same total baryonic number, 
the deconfinement transition is accompanied by the conversion of
predominantly two flavour matter
 into strange quark matter 
through weak interaction.
As a result, a large number of neutrinos may be   
produced during the phase transition from hadronic matter to strange 
quark matter. The annihilation of neutrinos \cite{gtr} thus produced 
gives rise to gamma ray bursts. Hence, we also estimate the duration 
of burst from the time span of neutrino production. 

In the following we proceed as follows. We first construct the full EOS
for the chemically equilibrated hadronic matter with phase transition to 
quark matter. This EOS gives us the final composition. 
{\it i.e.}, the fraction of baryons and quarks in the final star
in equilibrium. This in turn decides how much of the baryons are 
finally dissolved to quark matter. Once we know these fractions,
we can now use the formalism of ref.\cite{sannpa} to estimate
the total energy, number of neutrino and time taken for the
chemical equlibration during the conversion.

The full EOS of the neutron star with a quark core is constructed using two 
different models for hadronic and quark sector. The mixed phase is obtained 
following Gibbs criteria. The hadronic part of the EOS has been constructed 
using the TM1 parameter set of the non-linear Walecka model. The corresponding 
lagrangian density is \cite{a7}:
\begin{equation}
{\cal L} = {\cal L}_0 + {\cal L}_{YY} + {\cal L}_l
\end{equation}
where
\begin{eqnarray}
{\cal L}_0=\sum_B \psib_B\left(i\dd-m_B \right) \psi_B + {1 \over 2} 
\partial^\mu \sigma \partial_\mu \sigma - U \left(\sigma \right) 
-{1 \over 4} G^{\mu \nu} G_{\mu \nu} + U \left(\omega \right) 
-{1 \over 4} {\vec B}^{\mu \nu} {\vec B}_{\mu \nu} 
+ 
{1 \over 2} m_\rho^2 {\vec R}^\mu {\vec R}_\mu \nonumber\\
- \sum_B \psib_B \left(g_{\sigma B} \sigma + g_{\omega B}\omega^\mu\gamma_\mu 
+ g_\rho{\vec R}^\mu\gamma_\mu {\vec \tau}_B \right)\psi_B
\label{hadlag1}
\end{eqnarray}
\be
{\cal L}_{YY}={1 \over 2} \left(\partial^\mu \sigma^* \partial_\mu \sigma^* - 
- m^2_{\sigma^*} \sigma^{*2}\right)  - {1 \over 4} S^{\mu \nu} S_{\mu \nu} 
+{1 \over 2} m_\phi^2 \phi^\mu \phi_\mu 
- \sum_B \psib_B \left(g_{\sigma^* B} \sigma^* 
+ g_{\phi B} \phi^\mu\gamma_\mu \right)\psi_B
\ee
\be
{\cal L}_l=\sum_{l=e,\mu} \psib \left(i\dd-m_l \right) \psi_l 
\ee

In the above equations, $\psi_B$ is the baryon field and the $\sum_B$ runs 
over all the baryons ($p, n, \Lambda, \Sigma^0, \Sigma^+, \Sigma^-, 
\Xi^0$ and $\Xi^-$) and the $\sum_l$ runs over all the leptons. The piece 
of the Lagrangian ${\cal L}_{YY}$ is responsible for the hyperon-hyperon 
interactions \cite{a7}. The meson fields are 
$\sigma, \omega, {\vec R} ({\rho}), \sigma^* 
(f_0(975))$, and $\phi$. The $U_\sigma$ and $U_\omega$ are the $\sigma$ 
and $\omega$ meson potentials \cite{a7,a9,a10} which are given as : 
\be
U_\sigma = {b \over 3} \sigma^3 + {c \over 4} \sigma^4,  
U_\omega = {d \over 4} \omega^4 
\ee

As mentioned earlier, the TM1 parameter set has been used in this paper. 
The details of the parameter values may be obtained in ref.\cite{a5}. 
Quark matter EOS is obtained from the standard noninteracting MIT Bag model 
\cite{bag2}. Starting from the two models for the hadronic and the quark 
sectors, a first order deconfinement phase transition is obtained which 
proceeds via a mixed phase. At zero temperature, in the
presence of two conserved charges, the mixed phase is constructed following 
Gibbs criterion \cite{a3}. In the quark sector, we have taken the light quark 
masses to be zero, the strange quark mass to be 150 MeV and \( B^{1/4} \)= 
180 MeV. The EOS thus obtained has a mixed  
phase starting at about $3.8 \times 10^{14} g/cm^3$ and ends 
at about $1.76 \times  10^{15} g/cm^3$; for the sake of comparison, note 
that the energy density at nuclear saturation is 
$2.8 \times  10^{14} g/cm^3$.
The corresponding properties of the neutron star as well as the neutron star
with a quark core (hybrid star) for the maximum mass configuration are 
given in Table 1.

\begin{table}
\caption{Properties of non-rotating and rotating neutron as well as
hybrid stars (for maximum mass configuration); Rest mass (\( M_0/M_\odot \)),
gravitational mass ( \( M/M_\odot \) ), central energy density
( \( \epsilon_c \) ), equatorial radius ( \( R_e \) ) and
ratio of polar to equatorial radius (\(R_p/R_e \)). } 
\begin{tabular}{|c|c|c|c|c|c|c|}
Star motion&Star type & \( M_0/M_\odot \) &
 \( M/M_\odot \) & \( \epsilon_c \) & \( R_e \) & \( R_p/R_e \)  \\
\hline
Non-rotating& Neutron & 1.72 & 1.57 & \( 1.28 \times 10^{15} \)
& 13.63 & 1  \\
& Hybrid & 1.53 & 1.40 & \( 3.41 \times 10^{15} \)
& 10.20 & 1  \\
\hline
Rotating& Neutron & 2.12 & 1.93 & \( 1.10 \times 10^{15} \)
& 19.33 & 0.57  \\
(Mass shed limit)&  & & & 
&  &   \\
& Hybrid & 1.75 & 1.61 & \( 1.79 \times 10^{15} \)
& 16.77 & 0.59  \\
\end{tabular}
\end{table}
\begin{table}
\caption{Properties of stars corresponding to the EOS in fig. 1 and
fig. 2.}
\begin{tabular}{|c|c|c|c|c|}
\( \epsilon_c \) & \( M_0/M_\odot \) &
 \( M/M_\odot \) & Keplarian Frequency of & Keplarian Frequency of    \\
 &  &
 & the hybrid star (Hz)  & the initial Neutron Star (Hz)  \\
\hline
\( 6 \times 10^{14} g/cm^3 \) &  1.35 & 1.27 &  713 & 695  \\
\hline
\( 1 \times 10^{15} g/cm^3 \) &  1.62 & 1.50 &  895 & 767  \\
\end{tabular}
\end{table}
In case of rotating neutron stars, the energy density, and hence the
baryon density, profile is substantially different from that of a static 
star. The density profile can be obtained by solving Einstein's
equations using the full EOS. The metric and the 
procedures involved may be obtained from the ref. \cite{komatsu,bhat1}. 
The energy density profile of the star, rotating with
Keplarian frequency, for two different central densities,
\( 6 \times  10^{14} g/cm^3 \) (solid line) and \( 1 \times 10^{15} g/cm^3 \) 
(dashed line) are shown in figures 1 and 2. 

In figure 1, the energy density is plotted against the radial parameter s 
(integrated over angle \( \mu \equiv cos \theta \), \( \theta \) being the 
polar angle ), which
is defined as \( R/R_e = s/(1-s) \), \( R \) and \( R_e \) being the 
radius and the equatorial radius of the star respectively. Hence, at
equator \( R = R_e \) so that s=0.5. Figure 1 shows that a 
higher central energy density
results in a sharper variation in the profile. 

The \( \mu \) dependence of energy density (integrated over s) is shown in 
figure 2.  Here, the energy density is found to be much larger towards the polar
regions (smaller \( \mu \) ) for higher central energy density, where as,
the energy density towards the equator (higher \( \mu \)) is similar for
both the central energy densities. This has a strong bearing on the beaming 
angle of the emitted neutrinos, as we show later. The properties
of the compact star corresponding to
the two central energy densities (continuous and dashed curves 
in figs. 1 and 2) are given in table 2.
 
Let us now discuss the production of strangeness during the conversion 
of hadronic matter to SQM. The main reaction mechanism for the production 
of strange quarks within the quark matter is the non-leptonic weak interaction:  
\be
u + d \leftrightarrow u + s
\label{nonlep} 
\ee
As discussed earlier, the quark matter initially consists mainly of
u and d quarks. Thus, the chemical potentials of u and d are much larger
than that of s quark. The process in eqn.(\ref{nonlep}), converting d to s, 
releases energy, the amount of which depends on the difference between 
the d and s chemical potentials. Though reaction (\ref{nonlep}) is
the main agent for s production, the system is driven towards chemical 
equilibration mainly by the semi-leptonic weak interactions:
\ber
d(s) \rightarrow u + e^- + {\bar \nu_e}&;& 
u +e^- \rightarrow d(s) + \nu_e; \nonumber\\
d(s) + e^+ \rightarrow u +  {\bar \nu_e}&;& 
u  \rightarrow d(s) + e^+ + \nu_e;
\label{semilep}
\eer
The presence of positrons is mainly important for the cases where 
due to the trapping of neutrino and energy, the temperature of the
reaction region rises. In the present paper, we restrict 
ourselves to the case of a thin conversion front and assume that 
there is no trapping of neutrino and energy inside the star. Thus, the
star is at a constant temperature. The details of all the cases as well
as the behaviour of the reactions are given in ref.\cite{sannpa}. 
The semi-leptonic reactions are responsible for the neutrino production.
For chemically equilibrated matter, the rates of the reactions
given by eqn.(\ref{semilep}) are much smaller compared to those
for non-equilibrated matter \cite{sannpa}. 

The calculation proceeds as follows. Initial densities of quarks are obtained 
from the densities of different baryon species in the hadronic matter and 
their quark content. Final density fractions of the quarks are given by the 
density profile of the star as given in figures 1 and 2. Transition from the 
intial to the 
final state is governed by the rate equations:
\ber
{{dn_u(t)} \over {dt}} &=& R_{d \rightarrow u}(e^-) + R_{s \rightarrow u}(e^-) 
- R_{u \rightarrow d}(e^-) - R_{u \rightarrow s}(e^-) \nonumber \\
 &+& 
R_{d \rightarrow u}(e^+) + R_{s \rightarrow u}(e^+) - R_{u \rightarrow d}(e^+)
 - R_{u \rightarrow s}(e^+)  
\eer
where \( R_{d \rightarrow u}(e^-) \) is the reaction rate for the u quark
production from d quark via electron process. Other rates are defined
similarly. One can write down the rate equations for other quarks as well.  
Solving the coupled rate equations simultaneously along with the
chemical equilibrium conditions, we get the neutrino emission rate for 
different required density fractions. The total number of neutrinos produced 
during the transition can be obtained by folding this rate with the density 
profile of the star. Moreover, since the density profile depends on both the  
radial as well as the polar angles, integrating over radial coordinates gives us
the angular distribution of emitted neutrinos. If $n_\nu$ is the number 
of neutrinos emitted per unit time per baryon and $n_B$ is the baryon 
number density then the number of neutrinos emitted at a particular 
angle per unit time is given by 
\be
N_\nu = 2\pi \int r^2 dr \  n_\nu \ n_B {{e^{2\alpha+\beta}} \over {\sqrt{1-v^2}}}
\ee
where $\alpha$ and $\beta$ are the gravitational potentials and $v$ the 
rotational velocity. In figure 3 we have plotted the number of neutrinos 
as a function of 
$\mu$. One can see 
that, for central energy density \( 6 \times  10^{14} g/cm^3 \) (solid line),
there is a sharp peak between $\mu = 0.1$ and $\mu = 0.24$, the 
corresponding width being about $12^o$. If we increase the central energy 
density ( \( 1 \times 10^{15} g/cm^3 \), dashed line), matter concentration
towards the polar regions increases (figure 2) so that the angular variation
becomes smaller. This, on the other hand, means that for a star having 
higher central energy density and hence larger regions of 
quark matter, the beaming would be less pronounced.

The neutrino beaming found here may be the missing link that causes the
Gamma Ray Bursts. In general, the \( \nu \bar{\nu} \rightarrow e^+ e^- \) 
annihilation cross section is very small. But it has been shown \cite{gtr}
that the general relativistic effects may enhance this cross section
substantially and more than 10\% of the energy emitted in neutrinos
may be deposited in \( e^+ e^- \) pairs.  This enhancement is due to
the path bending of the neutrinos which in turn increases the probability 
of head on \( \nu \bar{\nu} \) collision.  Though a detailed
General Relativistic calculation is needed to quantify the resulting increment 
due to beaming, one can safely infer that beaming would increase the efficiency 
of the \( \nu \bar{\nu} \rightarrow e^+ e^- \) process further, providing a 
very efficient engine for the Gamma Ray Bursts. 

Our estimate shows that about \( 10^{52} \) ergs of energy is released in the 
form of neutrinos, the average energy of each neutrino being of the order of 
100 MeV. But some of the neutrinos may get trapped in the interior, thereby 
producing more \( \nu \bar{\nu} \) pairs. So the final number of neutrinos 
may somewhat be larger than the present estimate. Moreover, the trapping of 
neutrinos, and consequently energy, would result in an increase in temperature 
on the stellar interior. This, in turn, would result in further enhancement 
of the  \( \nu \bar{\nu} \rightarrow e^+ e^- \) cross section due to the 
gravitational red shift effect \cite{gtr}. 

Since the time required for the conversion of hadronic matter to 
quark matter at each density is calculable, one can estimate the 
time scale for the conversion of the neutron star to hybrid star 
or quark star. It has been shown earlier \cite{sannpa} that for 
a fixed temperature (say, T=10 MeV) the time taken to form the 
chemically equilibrated quark matter at density 0.6 \( fm^{-3} \) 
is around 0.1 s. This time would be smaller for lower density and/or 
higher temperature. In the present case the density at the core 
varies between 0.30 - 0.52 \( fm^{-3} \) and the corresponding time 
scale for chemical equilibration is \( 10^{-3} \) - \( 10^{-1} \).
Since Neutrinos are being emitted throughout during the conversion, 
this time scale would roughly be equal to the time duration of the 
Gamma Ray Burst. Hence, in our model, the duration of the Gamma Ray 
Burst is expected to be of the order of \( 10^{-3} \) - \( 10^{-1} \) 
seconds. Observationally, the duration of GRBs, 
range from \( 10^{-3} \) sec. to about \( 10^{3} \) sec. with a well 
defined bimodal distribution for bursts longer (long \& soft GRBs) or 
shorter (short \& hard GRBs) than t \( \approx \) 2 Sec. 
\cite{meszaros,kouveliotou}. So the present scenario would act as an engine 
for short \& hard GRBs.  
 
To conclude, we have shown that the rotation of the neutron star causes a
beaming of the neutrinos produced during the hadron to quark phase transition 
inside the star. This beaming, along with the general relativistic effects, 
can substantially enhance the neutrino annihilation cross section and thus 
provide a very efficient engine for Gamma Ray Bursts. The beaming angle 
depends on the extent of the quark phase inside the star. If the quark 
phase extent is smaller, then the neutrinos are emitted within a narrower 
angle and the corresponding time duration of the burst would be smaller. 
For the parameter values used here, the calculated angle, the emitted energy 
as well as the duration of the burst compare quite well with the observed 
values.

{\bf Acknowledgements}
AB would like to thank University Grants Commission, India for partial 
financial support through the grant PSW-083/03-04.

\newpage
\begin{figure}[t]
\vskip-2.1cm
\hskip-1.76cm
\centerline{\psfig{file=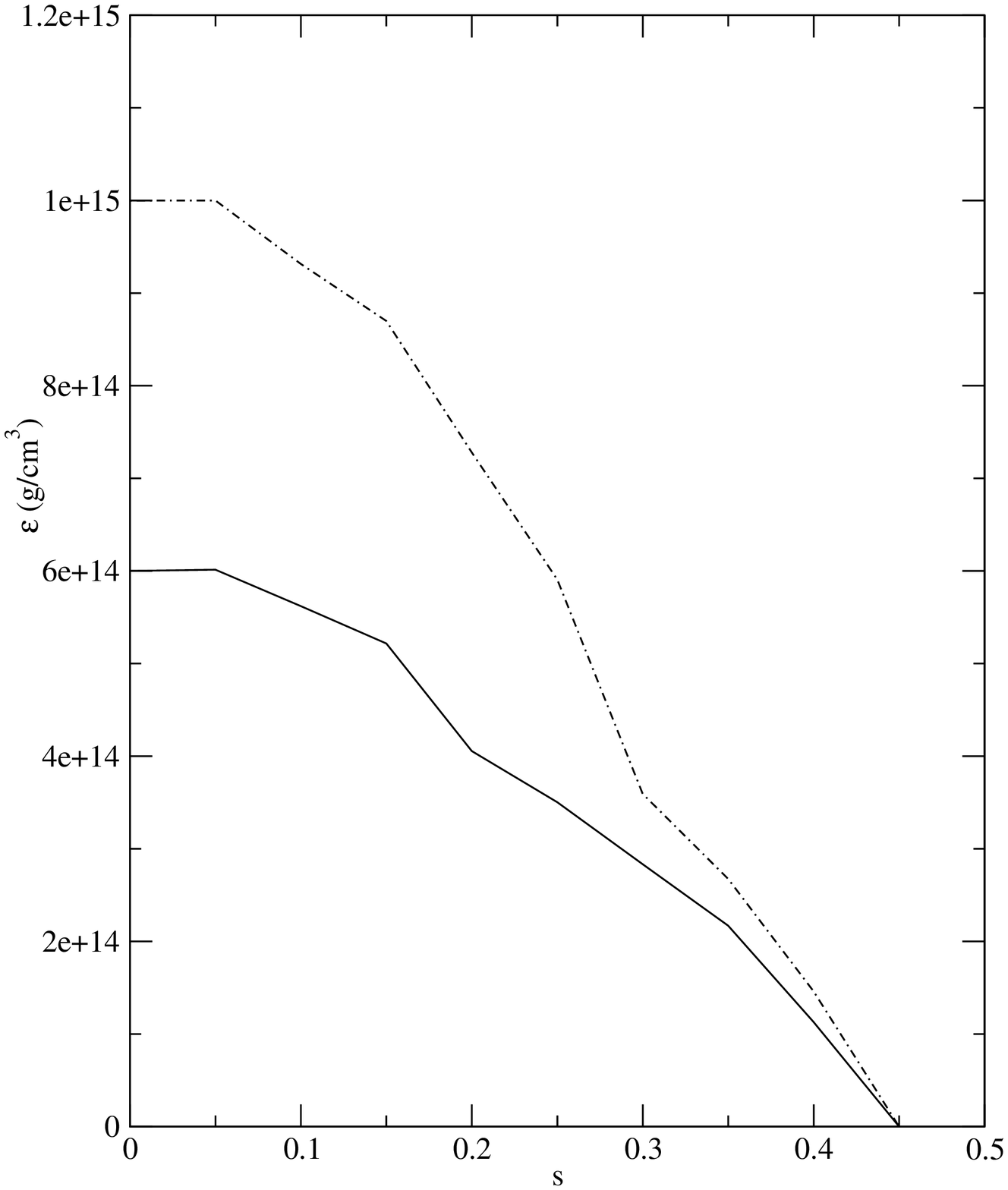,width=15cm}}
\caption{Energy Density variation (integrated over \( \mu \) ) with radial 
parameter s of the star for two central energy densities ; 
\( 6 \times 10^{14} g/cm^3 \) (solid line) 
and \( 1 \times 10^{15} g/cm^3 \)
(dashed line)}
\end{figure}
\newpage
\begin{figure}[t]
\vskip-2.1cm
\hskip-1.76cm
\centerline{\psfig{file=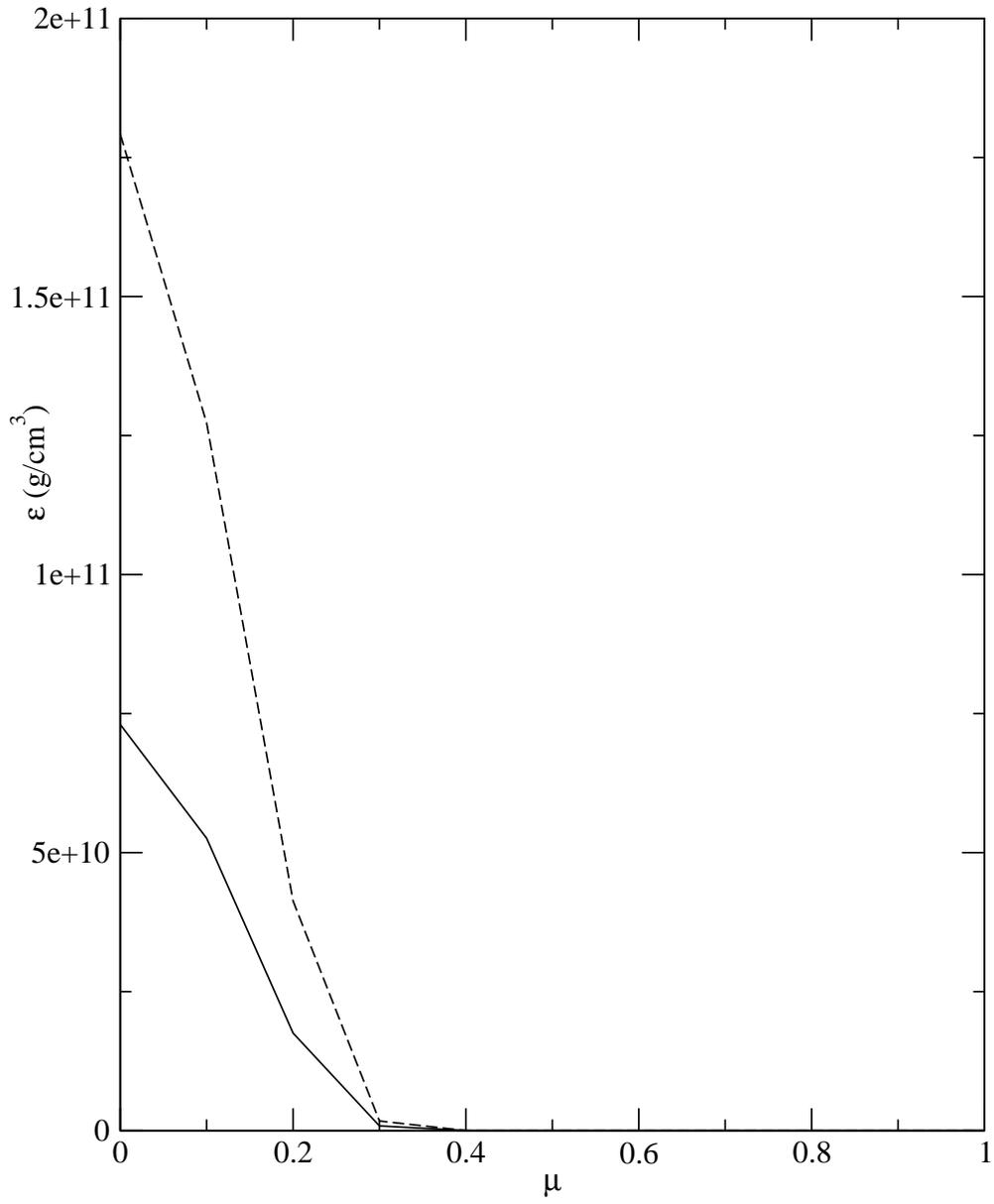,width=15cm}}
\caption{Angular distribution of the energy Density (integrated over radial 
parameter s) of the star for two central energy densities ; 
\( 6 \times 10^{14} g/cm^3 \) (solid line) 
and \( 1 \times 10^{15} g/cm^3 \)
(dashed line)}
\end{figure}
\newpage
\begin{figure}[t]
\vskip-2.1cm
\hskip-1.76cm
\centerline{\psfig{file=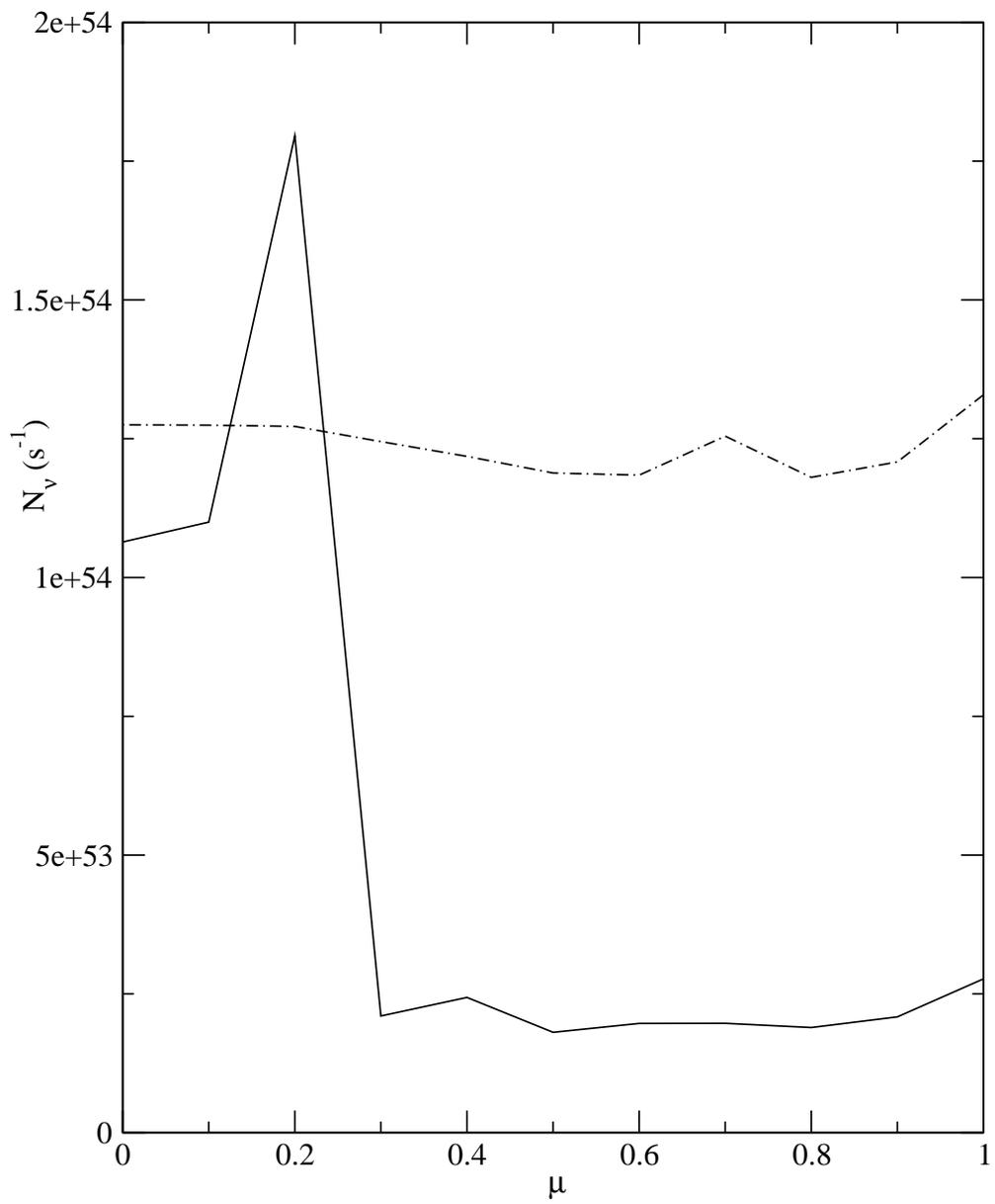,width=15cm}}
\caption{Neutrino emission as a function of $\mu$ for the central energy 
densities as in figs. 1 and 2.}
\end{figure}
\end{document}